\documentstyle[12pt]{article}
\newcommand{\be}{\begin{equation}}
\newcommand{\ee}{\end{equation}}

\newcommand{\lsim}{\mathrel{\mathop{\kern 0pt \rlap
  {\raise.2ex\hbox{$<$}}}
  \lower.9ex\hbox{\kern-.190em $\sim$}}}
\newcommand{\gsim}{\mathrel{\mathop{\kern 0pt \rlap
  {\raise.2ex\hbox{$>$}}}
  \lower.9ex\hbox{\kern-.190em $\sim$}}}
\begin{document}
\thispagestyle{empty}
\setcounter{page}{0}
\begin{flushright}
\begin{tabular}{r}
CERN-TH/96-283 
\\
DFTT 53/96
\\
October 1996
\end{tabular}
\end{flushright}  
\begin{center}
{\bf ON NEUTRALINO STARS AS MICROLENSING OBJECTS}\\
\vspace{5mm}
V.Berezinsky$^{a,c}$, A.Bottino$^b$ and G.Mignola$^{b,c}$\\
\vspace{5mm}
{\em $^a$INFN, Laboratori Nazionali del Gran Sasso, 67010 Assergi (AQ), Italy}\\
{\em $^b$Dipartimento di Fisica Teorica dell'Universit{\'a} di Torino, 
and INFN, Sezione di Torino, Via P. Giuria 1, I-10125 Turin, Italy} \\
{\em $^c$Theoretical Physics Division, CERN, CH-1211 Geneva 23, Switzerland}\\
\vspace{30mm}
{\large Abstract}
\end{center}
\vspace{3mm}
The microlensing objects, Machos, recently observed in the halo of our 
Galaxy, can be interpreted as dense neutralino objects,
neutralino stars, produced by the gravitational instability of neutralino 
gas. Taking the mass and radius of these objects from microlensing 
observations we calculated the diffuse gamma-ray flux produced in 
neutralino--neutralino annihilation inside the objects. 
The resulting flux is many orders of magnitude higher than the 
observed one. 
\vspace{10mm}
\begin{flushleft}
CERN-TH/96-283
\\
October 1996
\end{flushleft}

\newpage 

The recent observations of microlensing objects in the halo of our 
Galaxy \cite{macho} show that the Macho mass is
$0.46^{+0.30}_{-0.17}M_{\odot}$ at $68\%$ CL. This large mass of lenses raises
the question of how they can escape from the observations by the Hubble 
telescope. The fraction of the total halo mass in the  form of Machos, 
according to the observations \cite{macho}, is 
$\xi=0.50^{+0.30}_{-0.20}$. If this fraction is higher than $50\%$  a new 
question arises as to why the
fraction of non-baryonic cold dark matter, which is needed for the large scale 
structure formation, is so small in our Galaxy.   

In some recent papers \cite{Gur} a very interesting idea  
about the nature of these microlensing objects was put forward.
These objects can be dark matter formations around a singularity,
produced during the non-linear stage of the fluctuation growth (\cite{GZ} and 
references therein).
The authors call these objects neutralino stars (NS). If this idea is 
correct, the microlensing phenomenon is a natural result of fluctuation 
growth in the neutralino gas, and the aforementioned possible difficulties 
connected with the baryonic nature of Machos disappear. 

In this note we shall study the gamma-ray production in the neutralino 
stars in an approach more general than the model \cite{Gur}. Namely, we 
shall adopt for the neutralino stars the mass, the radius and their space 
density in the halo being restricted by the microlensing observations. For 
the density distribution inside NS we shall use $\rho \propto r^{-1.8}$, which 
is a  basic theoretical result for this kind of objects \cite{GZ}. We shall
 consider the mass of 
neutralino as a free parameter, though in the work \cite{Gur} the 
neutralino mass is restricted as $5~\mbox{GeV} \lsim m_{\chi} 
\lsim 50~\mbox{GeV}$.  

 The diffuse gamma-ray flux from neutralino stars  in the halo is  
\be
I_{\gamma}=\frac{1}{4 \pi}n_{s}R_h \dot{N}_{\gamma}\;,
\ee
where $n_{s}$ is the number density of neutralino stars in the halo,
$R_h$ is the radius of the halo and $\dot{N}_{\gamma}$ is the production rate 
of photons in the NS.

The density of NS in the halo can be estimated by assuming that the fraction
$\xi$ of the total mass of the halo, $M_h$, is in the form of NS: 
\be
n_{s}=\frac{3 \xi M_h}{4 \pi M_{\chi}R_h^3}
\ee
where $M_{\chi}$ is the mass of the NS.

Let us now estimate the production rate of photons $\dot{N}_{\gamma}$ in an NS.

The density of neutralinos inside the NS depends on the distance as $r^{-1.8}$ 
\cite{GZ} and can 
be expressed through the mass $M_{\chi}$ and radius $R_{\chi}$ of the NS as
\be
n_{\chi} \approx \frac{1.2}{4\pi}\frac{M_{\chi}}{m_{\chi}}R_{\chi}^{-3}
\left( \frac{r}{R_{\chi}} \right)^{-1.8}
\ee
with a negligible correction due to a modification of the $r^{-1.8}$ dependence 
at very small distances.
The gamma-ray production rate is
\be
\dot{N}_{\gamma}=2\eta_{\pi^0} 4\pi \langle \sigma v \rangle 
\int_{R_c}^{R_{\chi}}
n_{\chi}^2(r)r^2 dr
\ee
where $\langle \sigma v \rangle$ is 
the cross-section of neutralino--neutralino annihilation and 
$\eta_{\pi^0}(m_{\chi})$ is the 
multiplicity of neutral pion production in neutralino--neutralino annihilation.

We have assumed in (4) that the neutralino star has a core of radius 
$R_c$. Inside the core the neutralino density is constant or 
rising more slowly than $r^{-1.8}$, and we neglected the gamma-ray 
production there, which is 5 times smaller than outside it for the case of 
constant density.

Using Eqs. (1), (3) and (4) we obtain the diffuse flux as

\be
I_{\gamma}=\frac{0.9}{4\pi^3}\eta_{\pi^0} \frac{\xi M_h M_{\chi}}{m_{\chi}^2}
\left(\frac{R_{\chi}}{R_c}\right)^{0.6}
\frac{\langle \sigma v \rangle}{R_{\chi}^3 R_h^2}\;.
\label{eq:flux}
\ee

To proceed further we should evaluate two parameters: core radius $R_c$ and
annihilation cross-section $\langle \sigma v \rangle$.

 Let us start with the annihilation cross-section $\langle \sigma v \rangle$.
This quantity is of relevance both for the determination of the flux (see Eq.
(5)) and for the
evaluation of the neutralino relic abundance. In the usual expression 
$\langle \sigma v \rangle= a+bx$, where $x \equiv m_{\chi}/T$,
valid in the non-relativistic limit, $a$ takes contribution from
the s-wave only, whereas $b$ contains both s and p-wave contributions. 
The relic abundance is given
by 
\be
\Omega_{\chi} h^2 = 8.67\cdot 10^{-11}\frac{1}{N_f^{1/2}}\;,
\frac{\mbox{GeV}^{-2}}{ax_f+(1/2)bx_f^2}
\ee                    
where $N_f \approx 100$ is the number of degrees of freedom at the epoch of 
the neutralino decoupling and $x_f = m_{\chi}/T_f \simeq 1/20$, $T_f$ being the
temperature at the decoupling. Situations in which this approximation is not 
valid are discussed, for instance, in \cite{pole}.

Let us note that the annihilation cross section in Eq. (5) has to be evaluated
at the present temperature ($T_0$) and then 
for $x_0 \simeq 10^{-7}$. This implies
that $I_\gamma$ critically depends on the size of the $a$ term 
in $\langle \sigma v \rangle$;
any suppression effect in the $a$ term entails a large depletion in the diffuse
flux emitted by NS. As will be shown later, the flux $I_\gamma$ has the
tendency to largely (by many orders of magnitude) exceed the experimental
bound. Thus in what follows we adopt the conservative approach of analysing
with special care situations in which the term $a$ is significantly suppressed.

The evaluation of $\langle \sigma v \rangle$ requires the inclusion of 
the full set of annihilation final states ($f \bar{f}$ pairs, gauge-boson pairs,
Higgs-boson pairs and Higgs gauge-boson pairs), as well as the
complete set of Born diagrams ($Z$, Higgs, squark, neutralino and chargino 
exchanges). Furthermore one should include loop contributions, leading to gluon
final states, which may be relevant in annihilation processes at the present
temperature.  
Analytical expressions and numerical codes for the annihilation cross section 
are available (for a complete set of references, see \cite{Kam}).

As discussed above, the case favourable for neutralino stars as microlensing 
objects is the one in which the s-wave annihilation, described by $a$, 
is strongly suppressed in 
comparison with the p-wave contribution given by $b$. This occurs in the gaugino
and higgsino dominated states when the neutralino is lighter than the $W$-boson 
and the final states are fermionic pairs. In the case of strong 
s-wave suppression ($bx_f \gg a$) one obtains from (6):

\be
b \simeq 1.4 \cdot 10^{-35}\frac{0.2}{\Omega_{\chi}h^2} ~\mbox{cm}^2.
\ee
Our way of writing the numerical factors in the previous expression is  
motivated by the observation 
\cite{tel} that for all cosmologically successful models the cold dark matter
contribution to $\Omega$ is given by $\Omega_{CDM}h^2 = 0.2\pm 0.1$.

Turning now to the calculation of $\langle \sigma v \rangle$ at the 
present time, we shall evaluate $a$ using $b$ as given by Eq. (7) and the 
ratio $a/b$, which is almost free from particle-physics uncertainties 
following the procedure suggested in Ref. \cite{BBD}.
Let us first consider the case when $m_b < m_{\chi} < m_W$, where 
$m_b$ is the mass of the $b$-quark. The dominant contribution to the cross 
section is given in this case by $b\bar{b}$ final states. 

For the gaugino-like case the ratio of $a/b$ is given
\cite{BBD} by
\be
a/b=6.3\cdot 10^{-4} \left (\frac{m_b/5\;\;\mbox{GeV}} {m_{100}}\right)^{2}/
f(y)\;,
\label{eq:gaug}
\ee
where $m_{100}=m_{\chi}/100~\mbox{GeV}$. 
The function $f(y)$, where $y$ characterizes the neutralino 
composition, is given by \cite{BBD}
\be
f(y)=\frac{567-108y+1242y^2-12y^3+2023y^4}{(9-6y+5y^2)^2}
\ee
and is  bounded as
\be
8 \leq f(y) \leq 120.
\ee

For a higgsino-like neutralino, the $a/b$ ratio is \cite{BBD}:
\be
\frac{a}{b} \approx 
4.1\cdot10^{-6} \left(\frac{m_b/5\;\;\mbox{GeV}} {m_{100}}\right)^{2}.
\label{eq:higgs}
\ee
 Notice that the smallness of the 
$a/b$ ratio in both the gaugino- and higgsino-like neutralinos is due to the 
well known suppression effect in the s-wave annihilation channel (proportional
to the square of the mass of the final-state fermion) due to helicity 
properties \cite{goldb}. 

We recall that for the case of a neutralino of a mixed gaugino--higgsino
composition the Higgs exchange in the annihilation cross section  is usually 
important and the size of the $a/b$ ratio depends critically on the values of 
the Higgs masses. For instance, for a light $A$ Higgs boson 
($M_A \lsim 2 m_\chi$)
the $a$ term dominates, at decoupling,  over the $b$ term in the
annihilation cross section. This is due to the fact that, for the Higgs exchange
diagrams, also the $b$-term is proportional to $m_b^2$ because of the Yukawa
couplings. The case of s-wave suppression here can occur only at the pole
$M_A=2 m_{\chi}$ (see Ref. \cite{BBD}), but in this case $\Omega_{\chi}h^2$
is less than cosmologically needed. 

Therefore, to minimize  the gamma-ray production in the NS
we consider the  Higgsino-like neutralino with the $a/b$ ratio given 
by Eq. (\ref{eq:higgs}).
Then using Eq. (7) we obtain, for the cross-section: 
\be
\langle \sigma v \rangle \approx 1.7 \cdot 10^{-30}m_{100}^{-2}~
\mbox{cm}^3\mbox{s}^{-1},
\label{eq:cr-sec}
\ee
Note that this is the smallest annihilation cross-section we can obtain.
Since $b$ is fixed by $\Omega_{\chi}h^2$ (Eq. (7)), the value of 
the cross-section in Eq. (\ref{eq:cr-sec}) 
is determined by the suppression of the s-wave contribution.

For a neutralino heavier than the $W$-boson, the situation is different. Among 
many channels with weak gauge bosons and Higgses in the final states, there 
are those where the s-wave is not suppressed at all and those where it is 
strictly forbidden. Thus the average suppression is not strong. 

A discussion of the core radius  $R_c$ is in order now. We shall consider 
first the case of a pure neutralino star, with no baryon contamination.
The core radius can be estimated by equating the rate of neutralino 
accumulation in the core, $4\pi R_c^2 n_c u_r$, and that of neutralino 
annihilation there, $(4\pi/3) R_c^3 n_c^2 \langle \sigma v \rangle$, 
where $n_c$ is the
density of neutralinos in the core and $u_r$ is the bulk streaming velocity
of neutralinos towards the centre (we shall omit the index $r$ in the following
discussion): 
\begin{equation}
R_c = \frac {3 u(R_c)} {n_c \langle \sigma v \rangle} \;.
\label{eq:rc}
\end{equation}
To estimate $u$, let us consider the Euler and Poisson 
equations \cite{LL}, which determine the flow of the neutralino gas:
\be
\frac{\partial u}{\partial t}+ u\frac{\partial u}{\partial r}+
\frac{\partial \phi}{\partial r} = 0,
\label{eq:Eu}
\ee
\be
\triangle \phi= 4\pi G \rho,
\label{eq:Po}
\ee 
where $\phi$ is the gravitational potential, $\rho$ is the gas density and 
$G$ is the gravitational constant.

The solution (3) corresponds to the quasi-stationary regime, when 
$\partial u/\partial t$ can be neglected. Putting (3) into Eq. (\ref{eq:Po})
one finds 
\be
\frac{\partial \phi}{\partial r}= \frac{G M_{\chi}}{R_{\chi}^2}
\left(\frac{r}{R_{\chi}}\right)^{-0.8}.
\label{eq:pot}
\ee
Now it is easy to solve Eq.(\ref{eq:Eu}), which gives
\be
u(r)= \sqrt{5}(\frac{r}{R_{\chi}})^{0.1} u_{ff},
\label{eq:u}
\ee
where $u_{ff}$ is the free-fall velocity for an NS:
\be
u_{ff}=\sqrt{\frac{2 G M_{\chi}}{R_{\chi}}}\;.
\label{eq:ff}
\ee

Another way to estimate $u(r)$ is just to assume that it is everywhere 
the free-fall velocity relative to the mass inside the radius $r$. In this 
case we obtain
\be
u(r)= \sqrt{0.4} \left(\frac{r}{R_{\chi}}\right)^{0.1} u_{ff},
\ee
to be compared with Eq. (\ref{eq:u}).

Now by inserting Eq. (\ref{eq:u}) into Eq. (\ref{eq:rc}) one finds 
\be
R_c=28 R_{14}^{-0.665} m_{100}^{-3.33} M_{01}^{0.55}~\mbox{cm},
\ee
where $R_{14}$ is the radius of the neutralino star $R_{\chi}$ in units of
$10^{14}~\mbox{cm}$ and $M_{01}$ is $M_\chi$ in units of $0.1 M_\odot$. 

We shall further minimize the flux (\ref{eq:flux}) by using the following 
set of parameters :\\
$m_{\chi}=m_W=80~\mbox{GeV}$,
$ \eta_{\pi^0}=10$, taken from $e^+e^- \to \mbox{hadrons}$ data at 
$\sqrt s=80~\mbox{GeV}$,   $R_h^{max}=200~\mbox{kpc}$ and consequently
$M_h= 10^{12}M_{\odot}$, $\xi=0.1$,
$M_{\chi}^{min}=M_{macho}=0.1 M_{\odot}$, $R_{\chi}^{max}=1\cdot 10^{14}~
\mbox{cm}$,
$\langle \sigma v \rangle=2.6\cdot 10^{-30}~\mbox{cm}^3\mbox{s}^{-1}$ and 
$R_c=1\cdot 10^2~\mbox{cm}$.

The diffuse flux at $E_{\gamma} > 70~\mbox{MeV}$ is
\be
I_{\gamma}^{min}\simeq 1\cdot 10^7 (10^2~\mbox{cm}/R_c)^{0.6}~
\mbox{cm}^{-2}\mbox{s}^{-1}\mbox{sr}^{-1},
\label{eq:flux1}
\ee 
almost 12 orders of magnitude higher than allowed by observations of SAS 2 
and EGRET 
($I_\gamma \le 2 \cdot 10^{-5}~\mbox{cm}^{-2}\mbox{s}^{-1}\mbox{sr}^{-1}$ at 
$E_{\gamma} > 70~\mbox{MeV}$
\cite{gamma-exp}).
As was mentioned above the value of $I_\gamma$ in Eq. (\ref{eq:flux1}) is 
already a
conservative estimate in terms of neutralino composition. Even adopting the
very extreme phenomenological assumption that in the annihilation at the 
present time the $b$ term is  dominant (i.e. $a/b \lsim 10^{-7}$), 
the value of the gamma flux would only be reduced by a factor $\sim 40$, as
compared with the value in Eq. (\ref{eq:flux1}), and then it would still be 
largely above the experimental bound.

Why is the flux in Eq. (\ref{eq:flux1}) so large?
It is because we compressed the considerable fraction of the neutralino 
mass in the halo in the dense NS, taking the radius of 
neutralino star $R_{\chi} \leq 10^{14}~\mbox{cm}$ from the focusing condition
\cite{Gur}. It is instructive to compare the flux (5) with the case when
neutralinos are distributed homogeneously in the halo. The diffuse gamma-ray 
flux is
\be
I_{\gamma}^{hom}=\frac{1}{4\pi} \beta_{\gamma} R_h,
\ee
where  the rate of gamma-ray production $\beta_{\gamma}$ is
$$
\beta_{\gamma}=(n^{hom}_{\chi})^2 2 \eta_{\pi^0}\langle \sigma v \rangle
$$
and $n^{hom}_{\chi}$
is the homogeneous density of neutralinos in the halo.
The ratio of fluxes from NS and from a homogeneous distribution of neutralinos
is
\be
\frac{I_{\gamma}^{NS}}{I_{\gamma}^{hom}}=
0.8 \frac{M_{\chi}}{\xi M_h} \left(\frac{R_{\chi}}{R_c}\right)^{0.6}
\left(\frac{R_h}{R_{\chi}}\right)^3.
\ee
This ratio can be easily rewritten as 
\be
\frac{I_{\gamma}^{NS}}{I_{\gamma}^{hom}} \sim
\frac{<\rho_{\chi}>_{NS}}{\langle \rho_{\chi} \rangle _{hom}}
\left(\frac{R_{\chi}}{R_c}\right)^{0.6}\;,
\ee
where one can easily recognize the compression factor  discussed above
as the ratio of average neutralino density $\langle \rho_{\chi} \rangle$ 
in the NS and in the halo.  Numerically this ratio is $\sim 10^{17}$.\\*[5mm]

The flux given by Eq. (\ref{eq:flux1}) is obtained for a case of a pure 
NS. In reality the neutralino star should unavoidably have some 
baryonic contamination, because dissipative baryonic gas is streaming to 
gravitational potential minimum created by neutralino gas and is 
accumulated there. This phenomenon was clearly recognized  and considered by
Gurevich et al. (see \cite{GZ} and references therein). We would like to 
note here that heating of the baryonic gas due to neutralino--neutralino 
annihilation (which was not taken into account) is very important for 
the dynamics of the baryonic accretion in vicinity of a singularity.  
For the mass of the baryonic object $M_b$  the ratio 
$M_b/M_{\chi}\geq 10^{-3}$ was suggested in Ref. \cite{Gur}. 
We shall use $M_b=0.01 M_{\chi}$
\cite{Gur1}. As a simple estimate shows, for this mass the neutralino 
trajectories are modified by the baryonic core at distances 
$R_c \leq 0.03 R_{\chi}$. However,
it is questionable how the space density of neutralinos is changed. In 
Refs. \cite{Zel,Berch} it is shown that the density distribution of 
a non-dissipative gas with baryonic contamination is the same as for 
a non-dissipative gas. 
Even assuming that the core radius is very large,
$R_c=0.1 R_{\chi}$ \cite{Gur1}, the calculated flux
turns out to be  $I_{\gamma} \simeq 3 ~\mbox{cm}^{-2} 
\mbox{s}^{-1}\mbox{sr}^{-1}$, still 5 orders 
of magnitude larger than allowed by observations.

In conclusion, our analysis shows that the interpretation of the observed 
Machos or even of a small fraction of them as neutralino stars results in 
too large a diffuse gamma-ray flux, many orders of magnitudes higher than 
the observed one 
\footnote{
We do not analyse here the peculiar case of a very light photino considered in
Ref. \cite{Kolb}.
In this model, the neutralino relic abundance is determined by a physical 
process different from pair annihilation, and then the procedure presented in
our paper does not apply. The cross-section 
of photino--photino annihilation in this case can be small, since the large 
mass of sfermions and thus gamma-ray flux can be further suppressed.
However, we remark that in this model 
the photino mass is constrained within $\lsim 2\;\;\mbox{GeV}$ 
and then, below the 
range $5\;\mbox{GeV} \lsim m_\chi \lsim 50\; \mbox{GeV}$, allowed for the 
NS model of Ref.\cite{Gur}.}. \\*[10mm]  
\noindent
{\Large \bf Acknowledgements} \\
We are grateful to A.V. Gurevich for interesting and stimulating 
discussions.\\
V.B. thanks the TH Division of CERN for the hospitality.

\end{document}